\documentclass[aps,pre]{revtex4}

\usepackage{graphicx}
\usepackage{dcolumn}
\usepackage{amsmath}

\begin{document}

\title{Representational capacity of a set of independent neurons}
\author{In\'es Samengo}
\email{samengo@cab.cnea.gov.ar}

\author{Alessandro Treves}
\email{ale@sissa.it}
\affiliation{Programme in neuroscience -
S.I.S.S.A.\\ Via Beirut 2 - 4,
(34014) Trieste, Italy}

\date{01 Jannuary 2001}

\begin{abstract}
The capacity with which a system of independent neuron-like units represents
a given set of stimuli is studied by calculating the mutual information
between the stimuli and the neural responses. Both discrete noiseless
and continuous noisy neurons are analyzed. In both cases, the
information grows monotonically with the number of neurons considered.
Under the assumption that neurons are independent, the mutual
information rises linearly from zero, and approaches exponentially
its maximum value. We find the dependence of the initial
slope on the number of stimuli and on the sparseness of the representation.
\end{abstract}


\maketitle


\section{Introduction}

Neural systems have the capacity, among others, to represent stimuli,
objects and events in the outside world. Here, we use the word
{\it representation} to refer to an association between a certain
pattern of neural activity and some external correlate. Irrespectively of
the identity or the properties of the items to be represented, information
theory provides a framework where the capacity of a specific coding scheme
can be quantified. How much information can be extracted from the activity
of a population of neurons about the identity of the item that is being
represented at any one moment? Such a problem, in fact, has already been
studied experimentally
\cite{Optican,Eskandar,Kjaer,Heller,Tovee,Rolls1,Rolls3,Rolls2,Ale1,faces,libroale}.
Typically a discrete set of $p$ stimuli is presented
to a subject, while the activity of a population of $N$ neurons is recorded.
At its simplest, this activity can be described as an $N$ dimensional
vector ${\bf r}$, whose components are the firing rates of individual
neurons computed over a predefined time window. The measured response is
expected to be selective, at least to some degree, to each one of the stimuli.
This degree of selectivity can be quantified by the mutual information between
the set of stimuli and the responses \cite{Shannon}
\begin{equation}
I = \sum_{s = 1}^p P(s) \sum_{{\bf r}} P({\bf r} | s) \ \log_2
\left[ \frac{P({\bf r} | s)}{P({\bf r})} \right],
\label{inf}
\end{equation}
where $P(s)$ is the probability of showing stimulus $s$,  $P(
{\bf r} | s)$ is the conditional probability of observing response
${\bf r}$ when the stimulus $s$ is presented and
\begin{equation}
P({\bf r}) = \sum_{s = 1}^p P(s) \ P({\bf r} | s).
\label{pglob}
\end{equation}
The mutual information $I$ characterizes the mapping between the $p$
stimuli and the response space, and represents the amount of information
conveyed by ${\bf r}$ about which of the $p$ stimuli was shown. If
each stimulus evokes a unique set of responses, i.e. no two different
stimuli induce the same response, then Eq. (\ref{inf}) reduces
to the entropy of the stimulus set, and is, therefore, $\log_2 p$.
On the other hand, if a response ${\bf r}$ may be evoked by more than
one stimulus, the mutual information is less than the entropy of the stimuli.
In the extreme case where the responses are independent of the stimulus
shown, $I = 0$.

\begin{figure}
\begin{center}
\scalebox{.7}{\includegraphics{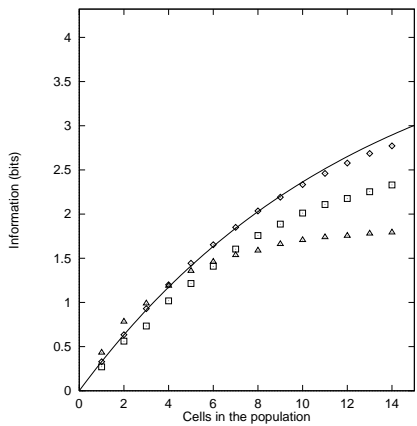}}
\end{center}
\caption{
Mutual information extracted from the activity of inferior temporal
cortical neurons of a macaque when exposed to $p$ visual stimuli.
Diamonds correspond to $p = 20$, squares
to $p = 9$ and triangles to $p = 4$. The graph is plotted as a
function of the number of neurons considered, once an average upon all
the possible permutations of neurons has been carried out.
The theoretical maximum is, in each case, $\log_2 p=4.32$ bits,
3.16 bits and, respectively, 2 bits. The full line shows a fit of Eq.
(\ref{grra}) to the case of $p = 20$.}
\label{f1}
\end{figure}

In Fig. \ref{f1}
we show the mutual information extracted from
neural responses from the inferior temporal cortex of a macaque when
exposed to $p$ visual stimuli \cite{Rolls2}. Diamonds correspond to
$p = 20$, squares to $p = 9$ and triangles to $p = 4$.
The graph is plotted as a function of the number of neurons considered.
Initially, the information rises linearly. As $N$ grows, the increase
of $I(N)$ slows
down, apparently saturating at some asymptotic value compatible with
the theoretical maximum $\log_2 p$.

The behavior shown in Fig. \ref{f1} is quite common observation also
in other experiments of the same type \cite{Rolls1,Rolls3,faces,libroale}.
From the theoretical point of view, different conclusions have been
drawn, over the years, from these curves.
Obviously, the saturation in itself implies that, after a while,
adding more and more neurons provides no more than redundant information.
Gawne and Richmond \cite{Gawne} have considered a simple model which
yields an analytical expression for $I(N)$ under the assumption that
each neuron provides a fixed amount of information, $I(1)$, and that a fixed
fraction of such an amount, $y$, is redundant with the information conveyed by
any other neuron. The model yields $I(\infty) = I(1)/y$.
Rolls {\it et. al} \cite{Rolls2} have considered a more constrained model that
in addition assumes that $y = I(1) / \log_2 p$. Later it was shown that this
is, in fact, the mean pairwise redundancy if the information provided by
different cells has a random overlap \cite{Samengo}. In this kind of
phenomenological description, the information provided by a population
of $N$ cells reads
\begin{equation}
I(N) = \log_2 (p) \left[1 - (1 - y)^N \right].
\label{grra}
\end{equation}
The full line in Fig. \ref{f1} shows a fit of Eq. (\ref{grra}) to
the data, in the case of $p = 20$.

It has also been suggested \cite{Rolls2} that monitoring the linear rise
for small $N$ may tell whether the representation of the stimuli is
distributed or local.
In a distributed scheme many neurons participate in coding for
each stimulus. On the contrary, in a local representation---sometiemes
called grandmother cell encoding---each stimulus is represented by the
activation of just one or a very small number of equivalent neurons.

Here we present a theoretical analysis of the dependence of $I$ on
$N$ for independent units. In contrast to the previous phenomenological
description, we model the response of  each neuron to every stimulus.
In Sects. II and III we derive $I(N)$ for several choices of the single
unit response probability. In Sect. IV we discuss the relation of the mutual
information defined in Eq. (\ref{inf}) to an informational measure of
retrieval accuracy.  We end in Sect. V with some concluding remarks.


\section{Discrete, noiseless units}

In what follows, the issue of quantifying the mean
amount of information provided by $N$ units is addressed. To do
so, the response of each unit to every stimulus is specified. From
such responses, the mutual information is calculated using Eq. (\ref{inf}).
Two types of models are considered. In this section we  deal with  discrete
noiseless units, while in Sect. III we turn to continuous noisy ones.

We consider $N$ units responding to a set of stimuli. The response $r_i$
of unit $i$ is taken to vary in a discrete set of $f$ possible values. The
states of the whole assembly of $N$ units are written as
${\bf r} \in {\cal R}$,
where ${\bf r} = (r_1, ..., r_N)$. Throughout the paper, letters in bold
stand for vectors in a $N$-dimensional space. The total number of states in
${\cal R}$ is therefore $f^N$.

The stimuli $\{s\}$ to be discriminated constitute a discrete set ${\cal S}$ of $p$
elements. For simplicity, we assume that they are all presented
to the neural system with the same frequency, namely
\begin{equation}
P(s) = \frac{1}{p}.
\label{stim}
\end{equation}
In order to calculate the mutual information between ${\cal S}$ and
${\cal R}$ we assume that each stimulus has a representation in ${\cal R}$.
In other words, for each stimulus $s$ there is a fixed
$N$-dimensional vector ${\bf r}^s$. Superscipts label stimuli,
while subscripts stand for units.

The fact that the neurons are noiseless means that the mapping between
stimuli and responses is deterministic. That is to say,
for every stimulus there is a unique response ${\bf r}^s$.
Mathematically,
\begin{equation}
P({\bf r} | s) = \left\{
\begin{array}{lll}
1 & {\rm if} & {\bf r} = {\bf r}^s, \\
0 & {\rm if} & {\bf r} \ne {\bf r}^s.
\end{array}
\right.
\label{noiseless}
\end{equation}
Therefore, for every $s \in {\cal S}$ there is one and only one ${\bf r} \in
{\cal R}$. The reciprocal, however, is in general not true. If
several stimuli happen to have the same representation---which may
well be the case if too few units are considered---then a given ${\bf r}$
may come as a response to more than one stimulus. In order to
provide a detailed description of the way the stimuli are associated
to the responses, we define $S_{{\bf r}}$ as the number of stimuli whose
representation is state ${\bf r}$. Clearly,
\begin{equation}
\sum_{{\bf r}} S_{{\bf r}} = p,
\label{sumk}
\end{equation}
and
\begin{equation}
P({\bf r}) = \frac{S_{{\bf r}}}{p}.
\end{equation}
When the conditional probability (\ref{noiseless}) is inserted in (\ref{inf}),
the sum on the responses can be carried out, since only a single vector
${\bf r} = {\bf r}^s$ gives a contribution. The mutual information reads
\begin{equation}
I = \sum_{{\bf r}} \frac{S_{{\bf r}}}{p} \ \log_2\left(\frac{p}{S_{{\bf r}}}
\right).
\label{idis}
\end{equation}
Thus, $I$ is entirely determined by the way the stimuli are clustered in the
response space. For example:

\begin{itemize}

\item
Consider the case where all stimuli evoke the same response. This
means that all the ${\bf r}^s$ coincide. Accordingly, $S_{{\bf r}^s} = p$
while all the other $S_{{\bf r}^\ell}$ vanish. There is no way the
responses can give information about the identity of the representations,
and $I = 0$.

\item
If every stimulus evokes its distinctive response there
are no two equal ${\bf r}^s$. This means that a number $p$ of the
$S_{{\bf r}}$ are equal to one, while the remaining vanish. The responses fully
characterize the stimuli, and $I = \log_2 p$.

\item
Consider the case of even clustering, where the representations
are evenly distributed among all the states of the system. This, or something
close to it, may in fact happen when the number of representations is much
larger than the number of states $p \gg f^N$.
Thus, $S_{{\bf r}} = p / f^N$, for all ${\bf r}$, and
$I = \log_2 (f^N)$. This is the maximum amount of information that can be
extracted when the set of stimuli has been partitioned in $f^N$ subsets,
and the responses are only capable of identifying the subsets, but not
individual stimuli.

\end{itemize}

\subsection{A local coding scheme}

We now consider another example, namely that of a local coding scheme,
sometimes called a system of {\it grandmother cells}. In 1972 Barlow
proposed a single neuron doctrine for perceptual psychology \cite{Barlow}.
If a system is organized in order to archieve as complete a representation
as possible with the minimum number of active neurons, at progressively
higher levels of sensory processing fewer and fewer cells should be
active. However the firing of each one of these high level units should code
for a very complex stimulus (as for example, one's grandmother). The
encoding of information of such a scheme is described as local.

Local coding schemes have been shown to have several drawbacks
\cite{libroale}, as their extreme fragility to the damage of the
participating units. Nevertheless, there are some examples in the
brain of rather local strategies such as, for example, retinal ganglion
cells (only activated by spots of light in a particular position
of the visual field \cite{Kuffler})
or the rodent's hippocampal place cells
(only responding when the animal is in a specific location in
its enviroment \cite{O'Keefe}).

We now evaluate the mutual information in such grandmother-cell
scheme, making use of Eq. (\ref{idis}).
For simplicity, we take the units to be binary ($f = 2$). We assume that
each unit $j$ responds to a single stimulus $s(j)$. Let us take that
response to be 1, and the response to any other stimulus to be 0.
All units are taken to respond to one single stimulus and, at first,
we take at most one responsive unit  per stimulus.
Thus, for the time being, $N \le p$.

\begin{figure}
\begin{center}
\rotatebox[origin=tr]{270}{\scalebox{.3}{\includegraphics{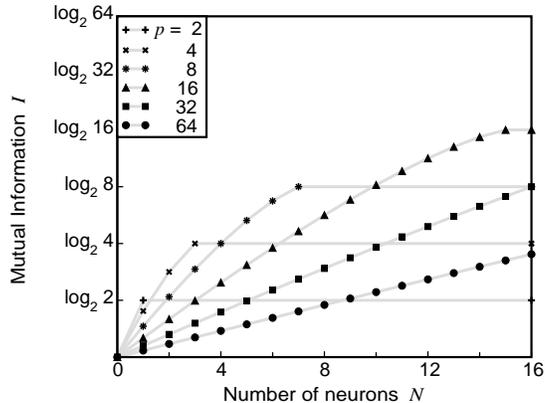}}}
\end{center}
\caption{Mutual information $I$ as a function of the number of
cells $N$, for different sizes of the set of stimuli, in the case
of localized encoding. For small $N$, the information rises
linearly with a slope proportional to $1/p$. When $N = p - 1$,
$I$ saturates at $\log_2 p$. } \label{f2}
\end{figure}

This particular choice for the representations means that out of the
$2^N$ states of the response space, only a subset of $N + 1$ vectors
is ever used. Actually, $S_{\bf 0} = p - N$, while for all
one-active-unit states ${\bf e}$, $S_{\bf e} = 1$. For the remaining
responses, $S_{{\bf r}} = 0$. Therefore, the mutual information reads
\begin{equation}
I = \frac{N}{p} \log_2\left(p\right) +
\frac{p - N}{p} \log_2\left(\frac{p}{p - N} \right).
\label{nonna}
\end{equation}
In Fig. \ref{f2}
we show the dependence of $I$ on the number of cells,
for several values of $p$. It can be readily seen that
for $N \ll p$
\begin{equation}
I \approx \frac{N}{p} \frac{1 + \ln p}{\ln 2} + {\cal O}(N / p)^2
\label{grnchic}
\end{equation}
In the limit of large $p$ Eq. (\ref{grnchic}) coincides with the intuitive
approximation
\begin{equation}
I(N) = N I(1) = N \left[ \frac{1}{p} \log_2 p + \frac{p - 1}{p}
\log_2\left(\frac{p}{p - 1} \right) \right].
\label{intuit}
\end{equation}

A linear rise in $I(N)$ means that different neurons provide different
information, or, in other words, that there is no redundancy in the
responses of the different cells. As seen in Fig. \ref{f2},
\begin{figure}
\begin{center}
\rotatebox[origin=tr]{270}{\scalebox{.3}{\includegraphics{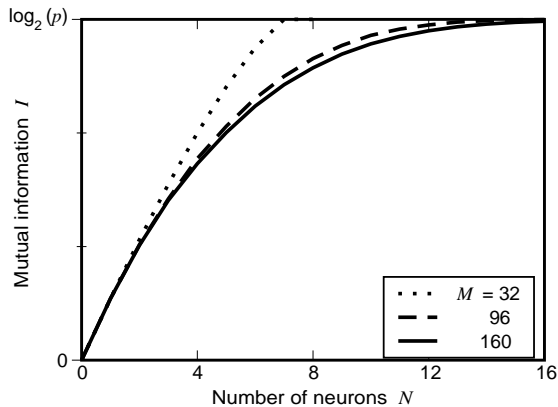}}}
\end{center}
\caption{ Mutual information $I$ as a function of the number of
recorded units $N$, once averaged over all the possible
selections of $N$ cells picked up from a pool of $M$ (the latter
constituted of $M / p$ units responding to each stimulus).
Different curves correspond to various values of $M$, and  $p =
32$.} \label{f2a}
\end{figure}
this is, in fact, the case when $N$ is small and $p$ is large
enough. When a cell does not respond, it is still providing some
information, namely, that it is not recognizing its specific
stimulus. When two cells are considered, a part of this
non-specific information overlaps with the information conveyed
by the second cell, when responding. In other words, if two cells
respond to different stimuli then, when one of them is in state
1, the other is, for sure, in state 0. Therefore, strictly
speaking, the information provided by different neurons in a
grandmotherlike encoding is not independent. However, in the
limit of $N/p \to 0$ the number of stimuli not evoking responses
in any single cell is large enough as to make the information
approximately additive.

As $N$ approaches $p$, such an independence no longer holds, so
the growth of $I(N)$ decelerates, and the curve
approaches $\log_2 p$. For $N = p - 1$, the mutual
information is exactly equal to $\log_2p$, and remains constant when
more units are added. In fact, $p - 1$ noiseless units are enough
to accurately identify $p$ stimuli. If all $p - 1$ are silent, then the
stimulus shown is the one represented by the missing unit.

In a slightly more sophisticated approach, each unit can have any number of responses
$f$. But as long as the conditional probability $P(r_j | s)$ is
the same for all those $s$ that are not $s(j)$, Eq. (\ref{nonna})
still holds.

It should be kept in mind that up to now
we have considered the optimal situation,
in that different units always respond to different stimuli. If several
cells respond to the same stimulus, a probabilistic approach is needed
since otherwise, the growth of $I(N)$ depends on the order in which
the units are taken. Averaging over all possible selections of $N$ cells
from a pool of $M$ units (the whole set is such that there are $M / p$
cells allocated to each stimulus) the result shown in Fig. \ref{f2a}
is obtained. We have taken $p = 32$, and different curves correspond to
various values of $M$. The probabilistic approach smoothes the sharp
behavior observed in Fig. \ref{f2}. Actually, the asymptote $\log_2 p$
can only be reached when there is certainty that there are $p - 1$ units
responding to different stimuli, that is, for $N = 1 + (p - 2) M / p$.
However, it is readily seen that with $M / p$ as large as 5, the curves
are already very near to the limit case of
$M / p \to \infty$.


\subsection{Distributed coding schemes}

As an alternative to the local coding scheme described above,
we now treat the case of distributed encoding, ranging from sparsely to fully
distributed. However, in doing so, we employ a different approach, namely, we
average the information upon the details of the representation.

Equation (\ref{idis}) implies that the amount of information that can be
extracted from the responses depends on the specific representations of the
$p$ stimuli. Since it is desirable to have a somewhat
more general result, we define an averaged mutual information $\langle I
\rangle$
\begin{equation}
\langle I \rangle = \sum_{{\bf r}^1, ... {\bf r}^p}
 P_0\left({\bf r}^{1},...,{\bf r}^{p} \right) \ I,
\label{iprom}
\end{equation}
where the mean is taken over a probability distribution $P_0({\bf r}^1,
..., {\bf r}^p)$ of having the representation in positions ${\bf r}^1,...,
{\bf r}^p$. This distribution, of course, is determined by the
coding scheme used by the system. By averaging the information we depart
from the experimental situation, where the recorded responses strongly
depend on the very specific set of stimuli chosen. But, in return,
the resulting information characterizes, more generaly, the way neurons encode
a certain type of stimuli, rather than the exact stimuli that have
actually been employed.

We write $P_0$ as a product of single distributions for each representation,
\begin{equation}
P_0 \left({\bf r}^1, ..., {\bf r}^p \right) =
\Pi_{s = 1}^p \ \ P_1({\bf r}^s).
\label{indepmem}
\end{equation}
This implies that the representation of one item does not bias the
probability distribution of the representation of any other. In this
sense, we can say that Eq. (\ref{indepmem}) assumes that
representations are independent from one another.

If in one particuar experiment the set of stimuli is large enough to effectively
sample $P_1({\bf r}^s)$ the averaged information will be close to the
experimental result.

We further assume that there is a probability distribution $\rho(r_j)$
that determines the frequency at which unit $j$ goes into state $r_j$
(or fires at rate $r_j$). If $\rho$ is strongly peaked at a particular
state---which can be always be taken as zero---the code is said
to be sparse. On the contrary, a flat $\rho$ gives rise to a fully distributed
coding scheme.

Finally, we assume that different units are independent. In other words,
we factorize the probability that a given stimulus is represented by
the state ${\bf r}$ as
\begin{equation}
P_1({\bf r}) = \Pi_{j = 1}^N \ \ \rho(r_j).
\end{equation}

In order to average the information (\ref{idis}) we need to
derive the probability that stimuli are clustered into any
possible set of $\{S_{\bf r}\}$. Such a probability
reads
\begin{equation}
P\left(\left\{S \right\} \right)  =
\left(
\begin{array}{c}
p \\
\left\{ S \right\}
\end{array}
\right) \
\Pi_{{\bf r}} \ \ \left[ P_1 \left({\bf r} \right)
\right]^{S_{{\bf r}}},
\end{equation}
where
\begin{equation}
\left(
\begin{array}{c}
p \\
\left\{ S \right\}
\end{array}
\right) =
\frac{p!}{\Pi_{{\bf r}} \ \ S_{{\bf r}}!}.
\label{multin}
\end{equation}
Therefore, the average mutual information may be written as
\begin{equation}
\langle I \rangle =
\sum_{\left\{S \right\}}
 P\left(\left\{S\right\}\right) \
I.
\label{esta}
\end{equation}
The summation runs over all sets $\{S\}$ such that
$\sum_{{\bf r}} S_{{\bf r}} = p$.
Replacing Eq. (\ref{idis}) in (\ref{esta}), we obtain
\begin{equation}
\langle I \rangle = \sum_{\left\{ S \right\}}
\left(
\begin{array}{c}
p \\
\left\{ S \right\}
\end{array}
\right)
\ \Pi_{{\bf r}} \  \  \left[P_1({\bf r})
\right]^{S_{{\bf r}}} \ \sum_{{\bf r}'}
\frac{S_{{\bf r}'}}{p} \, \log_2 \left(
\frac{p}{S_{{\bf r}'}} \right).
\end{equation}
Rearranging the summation so as to explicitly separate out
a single $S_{{\bf r}_j}$ one may write
\begin{equation}
\langle I \rangle =
\sum_{{\bf r}}
\sum_{S_{{\bf r}} = 1}^p \frac{p!}{S_{{\bf r}}!}
\left[P(S_{{\bf r}})\right]^{S_{{\bf r}}}
\ \frac{S_{{\bf r}}}{p} \log_2 \left( \frac{p}
{S_{{\bf r}}} \right)
\frac{1}{\left( p - S_{{\bf r}} \right)!}
\ A,
\end{equation}
where $A$ is the sum over all other $S$, namely
\begin{equation}
A =
\sum_{\left\{S_{{\bf r}' \ne {\bf r}} \right\}}
\left(
\begin{array}{c}
p - S_{{\bf r}} \\
\left\{ S_{{\bf r}' \ne {\bf r}} \right\}
\end{array}
\right)
\, \Pi_{{\bf r}'' \ne {\bf r}} \left[P_1\left({\bf r}''\right) \right]^
{S_{{\bf r}''}}  = \left[ 1 - P_1\left({\bf r} \right) \right]
^{p - S_{{\bf r}}}.
\end{equation}
Thus,
\begin{equation}
\langle I \rangle =
\sum_{{\bf r}} \sum_{S_{{\bf r}} = 1}^{p - 1}
\frac{(p - 1)!}{(S_{{\bf r}} - 1)![(p - 1) -
(S_{{\bf r}} - 1)]!}
\log_2 \left(\frac{p}
{S_{{\bf r}}} \right) \ \left[P_1\left(
{\bf r} \right) \right]^{S_{{\bf r}}} \
\left[1 - [P_1\left({\bf r} \right) \right]^
{p - S_{{\bf r}}}.
\label{oh}
\end{equation}

We now discuss two particular cases of Eq. (\ref{oh}).
First, we take the encoding to be fully distributed, namely $\rho(r_j) =
1 / f$. Therefore, $P_1({\bf r}_j) = 1 / f^N$. If this is
replaced in the previous expression, we obtain
\begin{equation}
\langle I \rangle_{{\rm dis}} = \left(1 - f^{-N} \right)^{p - 1}
\ \sum_{S = 0}^{p - 2} \frac{(p - 1)!}{S!(p - 1 - S)!} \
\left(f^N - 1 \right)^{-S} \
\log_2 \left( \frac{p}{S + 1} \right).
\end{equation}
It may be seen that the dependence of the information on $f$ and $N$ always
involves the combination $f^N$. This means that neither the number of
units, nor how many distinctive firing rates each unit has are relevant
in themselves. Only the total number of states matters.

\begin{figure}
\begin{center}
\rotatebox[origin=tr]{270}{\scalebox{.3}{\includegraphics{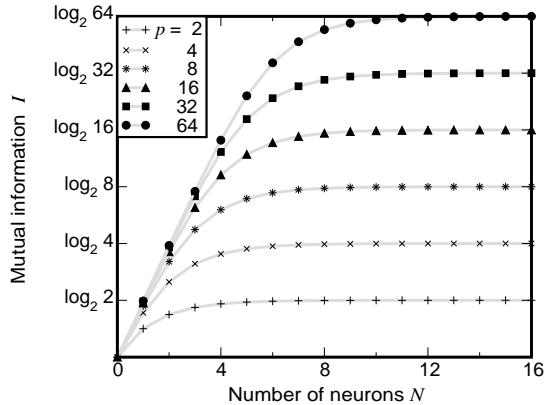}}}
\end{center}
\caption{ Mean mutual information  $\langle I \rangle_{ {\rm
dis}}$ as a function of the number of neurons $N$ for several
values of $p$. Initially the information rises linearly with a
slope only slightly depending on $p$. As $N$ increases, $\langle
I \rangle_{{\rm dis}}$ eventually saturates at $\log_2 p$. }
\label{f3}
\end{figure}

In Fig. \ref{f3}
we plot the relation between  $\langle I \rangle_{ {\rm dis}}$
and $N$ for several values of $p$. Initially the information
rises linearly with a slope only slightly dependent on $p$. As $N$
increases, $\langle I \rangle_{{\rm dis}}$ eventually saturates at
$\log_2 p$. The limit cases are easily derived
\begin{eqnarray}
\lim_{N \ln f \to 0}\langle I \rangle_{{\rm dis}} &=& N (p - 1) \ln f \log_2
\left(\frac{p}{p - 1} \right)  \label{chinpun} \\
\lim_{f^N / p \to \infty }\langle I \rangle_{{\rm dis}} &=& \log_2 p  - (p - 1) f^{- N}.
\end{eqnarray}
If the number of stimuli is large, Eq. (\ref{chinpun}) becomes
\begin{equation}
\lim_{N \ln f \to 0} \ \ \
\lim_{ p \to \infty }
\langle I \rangle_{{\rm dis}} = N \frac{p - 1}{p} \log_2 f.
\label{lindis}
\end{equation}
Notice that in contrast to the local coding scheme Eq. (\ref{nonna}),
the initial slope
of $I(N)$ hardly depends on $p$ (actually, it increases slightly with
$p$). This makes the distributed encoding a highly efficient way
to read out information about a large set of stimuli by the activity
of just a few units.

As opposed to the fully distributed case, a sparse distributed encoding
is now  considered, with $f = 2$, $\rho(1) = q$, $\rho(0) = 1 - q$ and
$q \ll 1$. This choice is again a binary case, but with one
response much more probable than the other. As a consequence,
the most likely representations in ${\cal R}$ space are those with either
zero or at most one active neuron. In fact,
$P_1({\bf r}^s = {\bf 0}) = (1 - q)^N$, whereas if the representation is a
one-active-unit state ${\bf e}$, $P_1({\bf e}) = q(1 - q)^{N - 1}$.
The probability of all other representations is higher order in $q$.

Accordingly, to first order in $q$, we only consider the combinations
of $p$ representations with at least $p - 1$ of them in state
${\bf r}^s = {\bf 0}$. These are the only responses
with a probability $P_0$ at most linear in $q$.
More precisely, the probability of representing all $p$ stimuli with the
same state ${\bf r} = {\bf 0}$ is
$P_0({\bf 0}, {\bf 0}, ..., {\bf 0}) = [P_1({\bf 0})]^p
\approx 1 - N p q$. In the same way, the probability of having
$N - 1$ stimuli in ${\bf 0}$ and a single one-active-unit
state is $q$. There are $N$ different possible one-active-unit
states, and any one of the $p$ stimuli can be such a state.
Taking all this into account, we find that up to the first order in
$N p q$,
\begin{equation}
\langle I \rangle_{{\rm spa}} = N p q \left[
\frac{p - 1}{p}
\log_2\left(\frac{p}{p - 1} \right) + \frac{1}{p} \log_2 p \right].
\label{iuju}
\end{equation}
Expanding this expression for large $p$, we obtain
\begin{equation}
\lim_{p \to \infty} \langle I \rangle = N q \frac{1 + \ln p}{\ln 2}.
\end{equation}
This means that from the experimental measurement of the slope of
$\langle I(N) \rangle$
it is possible to extract the sparseness of an equivalent binary model,
which can be compared with a direct measurement of the sparseness.
If the number of stimuli cannot be considered large, the whole
of Eq. (\ref{iuju}) can be used to derive a value for $q$.

It should be noticed that if $q = 1/p$ Eq. (\ref{iuju}) coincides with
the expression (\ref{intuit}) for a grandmother-like encoding.
This makes sense, since $q = 1/p$ implies that, on average, any
one unit is activated by a single pattern. In short, it corresponds to
a probabilistic description of the localized encoding. Notice, though,
that $q = 1/p$ is outside the range of validity of our limit $N p q \ll 1$.


\section{Continuous, noisy neurons}

In this section we turn to a more realistic description of the single neuron
responses. Specifically, we allow the states $r_j$ to take any real value.
Therefore, the response space ${\cal R}$ is now $\Re^N$. In
addition, we depart from the deterministic relationship between stimuli and
responses. This means that upon presentation of stimulus $s$, there is no
longer a unique response. Instead, the response vector ${\bf r}$
is most likely centered at a particular ${\bf r}^s$, and shows some
dispersion to nearby vectors. The aim is to calculate the mutual information
between the responses and the stimuli requiring as little as possible
from the conditional probability
$P({\bf r} | s)$. A single parameter $\sigma$
is introduced as a measure of the noise in the representation. Thus,
\begin{equation}
P\left({\bf r} | s \right) =
\Pi_{j = 1}^N \ \ \
\frac{{\rm e}^{-(r_j - r_j^s)^2/2\sigma^2}}{\sqrt{2 \pi \sigma^2}},
\label{ret}
\end{equation}
where the index $s$ takes values from $1$ to $p$. The conditional probability
depends on the distance between the actual response ${\bf r}$ and a fixed vector
${\bf r}^s \, \in {\cal R}$, which is the mean response
of the system to stimulus $s$. There is one such ${\bf r}^s$ for
every element in ${\cal S}$. The choice of Gaussian functions is only to keep
the description simple and analytically tractable. By factorizing
$P\left({\bf r} | s \right)$ in a product of one component probabilities
an explicit assumption about the independence of the neurons is being made.

Figure \ref{f4} shows  a numerical evaluation of the
information (\ref{inf}), when the probability $P({\bf r} | s)$ is
as in (\ref{ret}). The information, just as in the previous section,
has been averaged upon many selections of the representations ${\bf r}^s$.
The curve is a function of the number of neurons
considered $N$. Different lines correspond to different sizes of
the set of stimuli, while in (a) $\sigma = \lambda / 2$, and in (b)
$\sigma = \lambda$, where $\lambda$ is a parameter quantifying the mean
discriminability among representations, to be defined precisely later.
Just as in the discrete distributed case, we observe an initial
linear rise and a saturation at $\log_2 p$. Moreover, and pretty
much as in the experimental situation of Fig. \ref{f1},
the initial slope does not seem to depend strongly on the number of
stimuli, at least for large values of the noise $\sigma$.
In what follows, an analytical study of these numerical results is
carried out. In particular, the relevant parameters determining
the shape of $I(N)$ are identified.

We write the mutual information as
\begin{equation}
I = H_1 - H_2,
\label{separ}
\end{equation}
where
\begin{equation}
H_1 = - \frac{1}{p} \sum_{s = 1}^p \int d {\bf r}
P\left({\bf r} | s \right) \
\log_2 \left[\frac{1}{p} \sum_{s' = 1}^p
P\left({\bf r}|s' \right)
\right] =
 -  \int d {\bf r} \ P\left({\bf r} \right) \
\log _2 \left[P\left({\bf r}\right) \right],
\label{i2}
\end{equation}
is the total entropy of the responses, and
\begin{equation}
H_2 = - \frac{1}{p} \sum_{s = 1}^p \int d {\bf r}
P\left({\bf r} | s \right) \
\log_2 \left[
P\left({\bf r} | s \right)
\right].
\label{i1}
\end{equation}
is the conditional entropy of
$P\left({\bf r} | s \right)$, averaged over $s$.

$H_2$ can be easily calculated. It reads
\begin{equation}
H_2 = \frac{N}{2 \ \ln 2} \left[ 1 + \ln(2 \pi \sigma^2) \right].
\label{i1a}
\end{equation}
It is therefore linear in $N$. This stems from the
independence of the units, since the entropy of the response space
increases linearly with its dimension. It does not depend on
the location of the representations ${\bf r}^s$, and it is a growing
function of the noise $\sigma$.

In Appendix A we solve the integral in ${\bf r}$ of $H_1$ using the
replica method. We obtain
\begin{eqnarray}
H_1 &=& \frac{-1}{\ln 2}
\lim_{n \to 0}
\frac{1}{n}
\left(\left\{
\frac{1}{p^{n + 1} (2 \pi \sigma^2)^{N n /2}(n + 1)^{N / 2}}
\sum_{\{K\}}
\left(
\begin{array}{c}
n + 1 \\
\{K\}
\end{array}
\right)
\right. \right. \label{i2fin} \\
&\times& \left. \left.
\Pi_{j = 1}^N \ \exp\left[\frac{-1}{4 \sigma^2 (n + 1)}
\sum_{\ell = 1}^p \sum_{m = 1}^p
K_\ell K_m \left(r^\ell_j - r^m_j \right)^2
\right] \right\} - 1 \right), \nonumber
\end{eqnarray}
where $\{K\}$ now stands for the set $\{K_1, K_2, ... K_p\}$
specifying how many replicas are representing each pattern.
The summation in $\{K\}$ runs over all sets of $K$ such that
$\sum_{s = 1}^p K_s = n + 1$.
The symbol in brackets is defined in (\ref{multin}).
Equation (\ref{i2fin}) shows that the information depends explicitly on the
ratio between all the possible differences $|{\bf r}^\ell - {\bf r}^m|$
and the noise $\sigma$. In other words, the capacity to determine which
stimulus is being shown is given by a signal-to-noise ratio, characterizing
the discriminability of the responses.

The mutual information $I$ characterizes the selectivity of the
correspondence between stimuli and responses. If the distance between any two vectors
$|{\bf r}^\ell - {\bf r}^m|$ is much greater than the
noise $\sigma$, then the mapping is (almost) injective. Thus, in this limit
the mutual information approaches its maximal value, $\log_2 p$.

If, on the other hand, the noise level in $P\left({\bf r} | s\right)$
is enough to allow for some vectors ${\bf r}$ to be evoked with
appreciable probability by more than one stimulus, the mutual information
decreases. In this sense, $I$ can be interpreted as a comparison
between the noise in $P\left({\bf r} | s \right)$ and the
distance between any two mean responses. For a specific choice of the
representations,
the distance between any two of them is a non linear function of
their components. Therefore, in general, even though Eq. (\ref{ret})
implies that different units are independent, it is not possible
to write $I$ as a sum over units of single-units information.

Just as before, we now average the mutual information (\ref{inf}) over a
probability distribution $P_0({\bf r}^1, ..., {\bf r}^p)$ of
the representations ${\bf r}^1,..., {\bf r}^p$, namely
\begin{equation}
\langle I \rangle = \int \Pi_{j = 1}^p \ \   d{\bf r}^j
 \ P_0\left({\bf r}^{\,1},...,{\bf r}^{\,p} \right) \ I.
\label{iprom1}
\end{equation}
Under the assumption that the responses to different stimuli
are independent,  $P_0$ reads
\begin{equation}
P_0\left({\bf r}^{\,1}, ..., {\bf r}^{\,p} \right) = \Pi_{s = 1}^p
 \ \ P_1\left({\bf r}^{\,s} \right).
\label{patind}
\end{equation}
Adding the requirement of independent units,
\begin{equation}
P_1\left({\bf r}^s\right) = \Pi_{j = 1}^N \ \ \rho\left(
r^s_j \right).
\label{neuind}
\end{equation}
By replacing the average (\ref{iprom}) in the separation (\ref{separ})
we write
\begin{equation}
\langle I \rangle = \langle H_1 \rangle - H_2,
\label{i2e}
\end{equation}
since $H_2$ does not depend on the vectors ${\bf r}^s$.

So we now turn to the calculation of $\langle H_1 \rangle$, namely
\begin{eqnarray}
\langle H_1 \rangle = &-& \frac{1}{\ln 2}
\label{i2d} \\
& & \lim_{n \to 0} \frac{1}{n}
\left[
\frac{1}{(n+1)^{N/2} (2 \pi \sigma^2)^{N n / 2} p^{n + 1}}
\ \sum_{\{K\}}
\left(
\begin{array}{c}
n + 1  \\
\left\{K \right\}
\end{array}
\right)
\langle A_{\{K\}} \rangle^N - 1 \right],
\nonumber
\end{eqnarray}
where
\begin{equation}
\langle A_{\{K\}} \rangle = \int  \Pi_{s = 1}^p
d r^s \ \rho\left( r^s \right)  \
\exp \left[-\frac{1}{4(n + 1)\sigma^2} \sum_{m = 1}^p
\sum_{\ell = 1}^p K_m K_\ell (r^m - r^\ell)^2\right].
\label{i2c}
\end{equation}
The main step forward introduced by the average in (\ref{iprom}) is that
now, $\langle H_1 \rangle$ is symmetric under the exchange of any
two responses, or any two neurons. In contrast, before the averaging
process, the location of every single response by every single unit was
relevant.

The limit in Eq. (\ref{i2d}) can be calculated in some particular cases.
In the first place, we analyze the large $N$ limit. From Eq. (\ref{i2c}) it is
clear that $\langle A_{\{K\}} \rangle \le 1$. The equality holds, in fact,
only when there is a single $K$ different from zero. In the calculation
of $\langle H_1 \rangle$, as stated in Eq. (\ref{i2d}), $A_{\{K\}}$
appears to the $N$-th power. Therefore, when $N \to \infty$
only the terms with $A_{\{K\}} = 1$ give a non-vanishing contribution.
There are $p$ of such terms. When the sum in (\ref{i2d}) is replaced
by $p$, it may be shown that once more, $\langle I \rangle = \log_2 p$.

In the following two subsections we compute $\langle I(N) \rangle$
for both large and small values of the noise $\sigma$.


\subsection{Information in the large noise limit}

We now make the assumption that the noise $\sigma$ is
much larger than some average width of $\rho\left(r\right)$. In other
words, we suppose $\sigma^2 \gg (r^\ell - r^m)^2$, for all $r^\ell$
and $r^m$ with non-vanishing probability. In this case, the exponential
in (\ref{i2c}) may be expanded in Taylor series. Up to the second
order,
\begin{eqnarray}
\exp  \left[- \frac{1}{4(n + 1)\sigma^2} \sum_{m = 1}^p
\sum_{\ell = 1}^p K_m K_\ell (r^m - r^\ell)^2\right]
&\approx& 1 - \label{smals} \\
& & \frac{1}{4(n + 1)\sigma^2} \sum_{m = 1}^p
\sum_{\ell = 1}^p K_m K_\ell (r^m - r^\ell)^2
+ \nonumber \\
& & \frac{1}{2}\left[\frac{1}{4 (n + 1)\sigma^2} \sum_{m = 1}^p
\sum_{\ell = 1}^p K_m K_\ell (r^m - r^\ell)^2 \right]^2
\nonumber
\end{eqnarray}
If only the constant term is considered, the integral in Eq. (\ref{i2c})
becomes the normalization condition for $P_0$.  Thus,  the sums in (\ref{i2d})
give $p^{n + 1}$, and it is readily seen that $\langle H_1 \rangle$
exactly cancels $H_2$. As expected, in the limit
$\sigma^2 \to \infty$ the mutual information vanishes.

\begin{figure}
\begin{center}
\rotatebox[origin=tr]{270}{\scalebox{.2}{\includegraphics{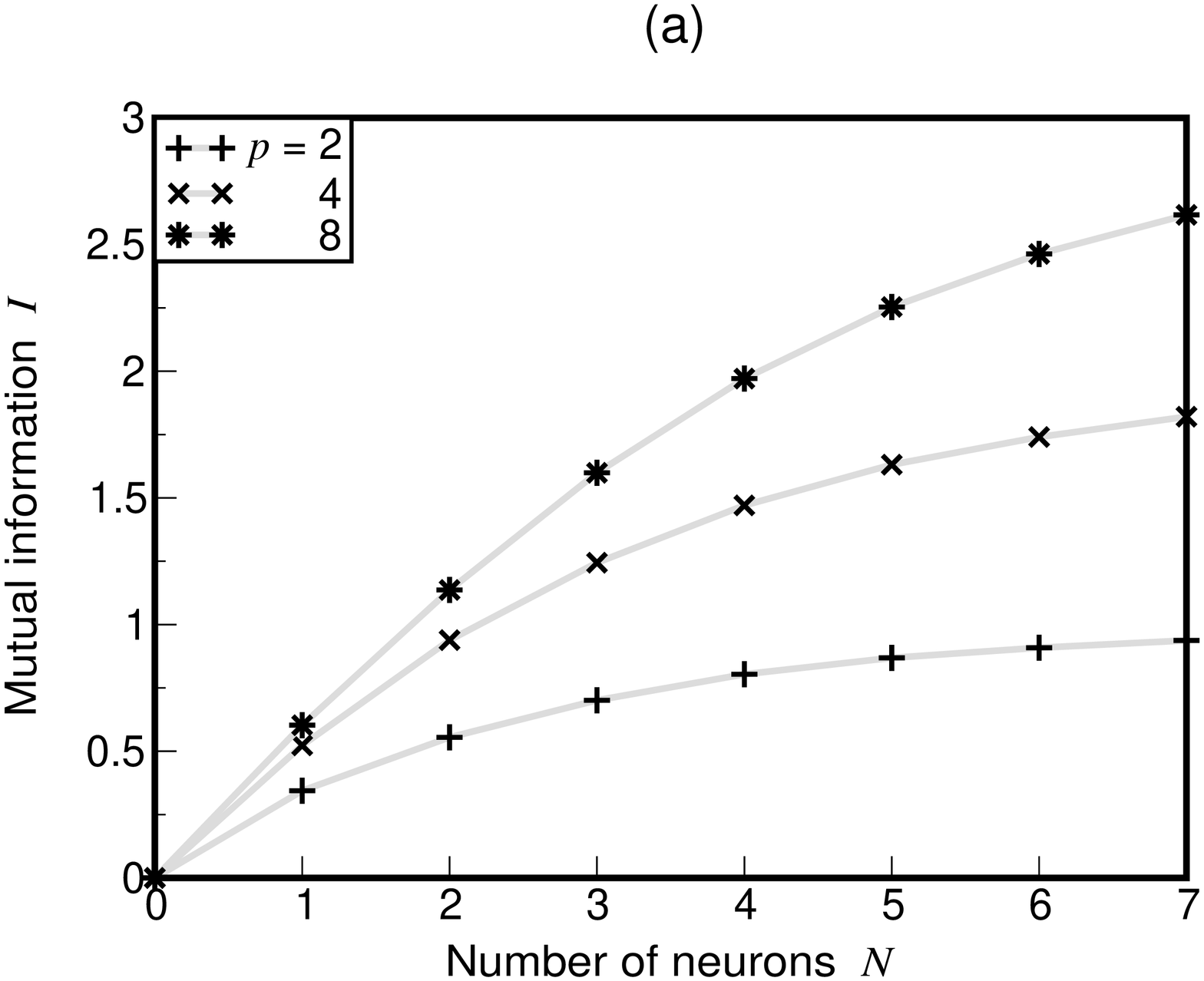}}}
\end{center}
\begin{center}
\rotatebox[origin=tr]{270}{\scalebox{.2}{\includegraphics{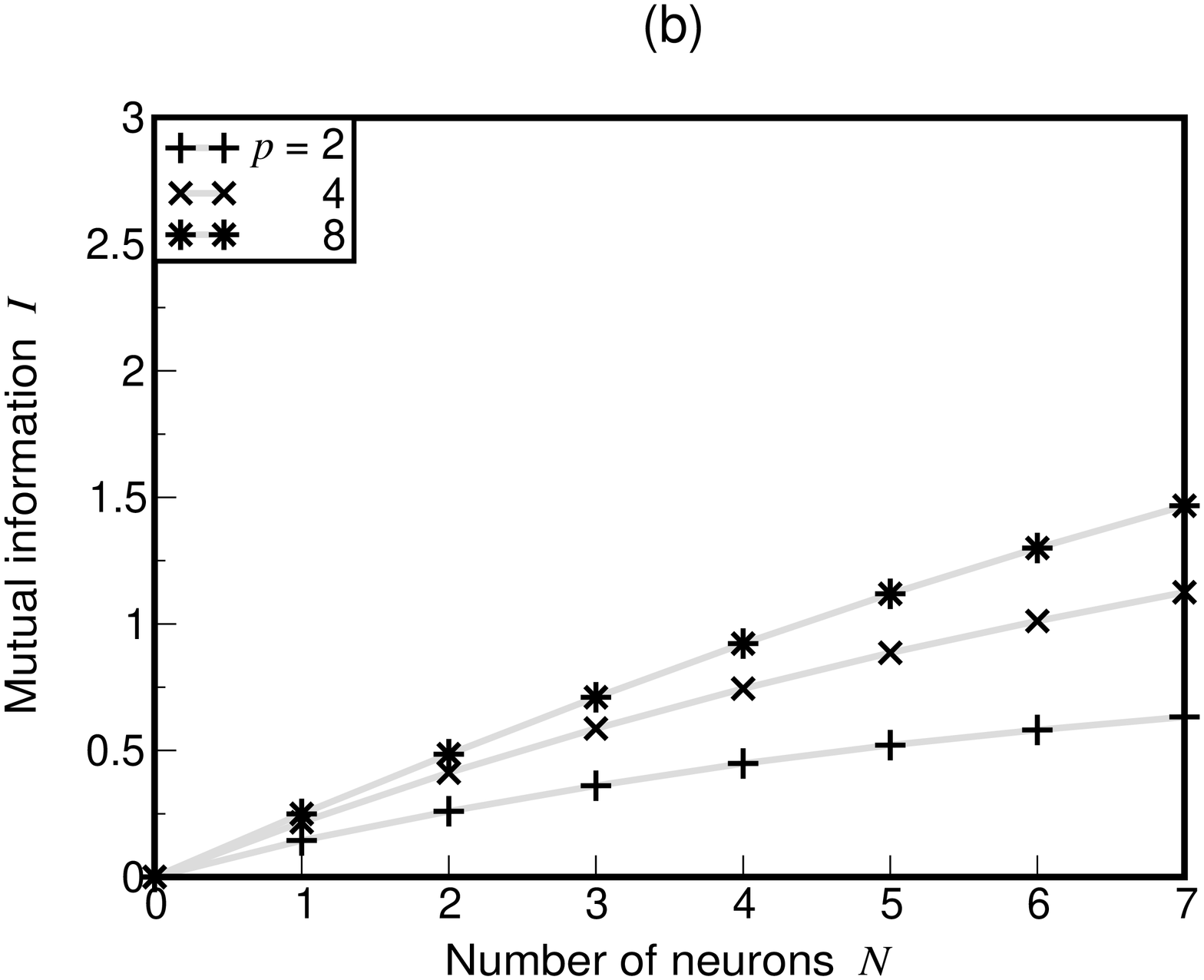}}}
\end{center}
\caption{ Results of the numerical evaluation of the mutual
information for continuous noisy neurons, where $p$ is the number
of stimuli in the set. In (a) $ \sigma = \lambda / 2$, and in (b)
$\sigma = \lambda$.} \label{f4}
\end{figure}

The next order of approximation is to consider the expansion
(\ref{smals}) up to the linear term.
Thus, the integral in (\ref{i2c}) becomes
\begin{equation}
\langle A_{\{K\}} \rangle = 1 - \frac{\lambda^2}{4(n + 1)\sigma^2}
\sum_{m = 1}^p \ \ \ \sum_{
\ell = 1,
\ell \ne m
}^p
K_m \, K_\ell,
\label{i2ca}
\end{equation}
where
\begin{equation}
\lambda^2 = \int d r^1 \ d r^2 \ \rho\left( r^1 \right)
\ \rho\left( r^2 \right) \ \left( r^1 - r^2 \right)^2
\end{equation}
is the parameter quantifying the discriminability among representations,
and appearing in Fig. \ref{f4}.
We have now gained a more precise insight of the large $\sigma$
limit. It stands for taking $\sigma \gg \lambda$.

Since in Eq. (\ref{i2d}) $A_{\{K\}}$ appears to the $N$-th
power, in order to proceed further we have to estimate the size of
$N \lambda^2 / \sigma^2$.
We first consider the small $N$ limit and assume, to start with,
that $N \lambda^2 / \sigma^2
\ll 1$. Thus, we may expand
\begin{equation}
\left( A_{\{K\}} \right)^N \approx 1 - \frac{N \lambda^2}{4(n +
1)\sigma^2} \sum_{m = 1}^p \ \ \ \sum_{ \ell = 1, \ell \ne m }^p
K_m \, K_\ell. \label{i2cd}
\end{equation}
In Appendix B we calculate the sums in (\ref{i2cd}), thus
obtaining $\langle H_1 \rangle$. When the result is replaced in
(\ref{i2d}) we get
\begin{equation}
\langle I \rangle = \frac{N}{\ln 2} \ \frac{(p - 1)}{p}
\ \left(\frac{\lambda}{2 \sigma}\right)^2.
\label{imchico}
\end{equation}
For a large amount of noise, the information rises linearly
with the number of neurons. This dependence should be compared with
Eq. (\ref{lindis}), in the discrete distributed case. The two expressions
coincide, if the number of discrete states $f$ is associated to
$\exp(\lambda^2 / 4 \sigma^2)$. Therefore, as regards to the mutual information,
a dispersion $\sigma$ in the representation is equivalent to having a number
$\exp(\lambda^2/4\sigma^2)$ of distinguishable discrete responses.
Notice that both the noisy, continuous  and the discrete,
deterministic approch show the same dependence on the number of
representations.

Regarding the dependence on $\sigma$, it is readily seen that as the noise
decreases, the slope of $I$ increases. In other words, every single neuron
provides a larger amount of
information. Since the mutual information saturates at $\log_2 p$ for
$N \to \infty$, a small value of $\sigma$ implies that the ceiling is
quickly reached. As a consequence, the assumption $N \ll \sigma^2/\lambda^2$
can now be more precisely stated as $N \ll (\sigma^2 / \lambda^2) \log_2 p$.
In this regime, linearity holds.

As $N$ increases, saturation effects become evident, and the mutual
information is no longer linear. The first hint of the presence
of an asymptote at $\log_2 p$ is given by the quadratic contribution to
$I(N)$. In order to describe it, the whole of expansion
(\ref{smals}) must be replaced in Eq. (\ref{i2c}). Carrying out the
integral in $r^1, ..., r^p$,
\begin{equation}
\langle A_{\{K\}} \rangle = 1 - \frac{\lambda^2}{4 \sigma^2 (n + 1)}
\sum_{\ell = 1}^p \sum_{
m = 1,
m \ne \ell
}^p
K_\ell K_m + \frac{\eta^4}{32 \sigma^4 (n + 1)^2},
\end{equation}
where
\begin{equation}
\eta^4 = \int \left[ \sum_{\ell = 1}^p \sum_{m = 1}^p
(r^\ell - r^m)^2 K_\ell K_m \right]^2 \
\Pi_{s = 1}^p \ \ \rho\left(r^s \right) \, dr^s
\label{eta}
\end{equation}
Extracting the sums from the integral, the limit in Eq. (\ref{i2d})
can be solved, and,
\begin{equation}
\langle I \rangle = \frac{N}{\ln 2} \frac{p - 1}{p} \left[
\frac{\lambda^2}{4 \sigma^2} + \frac{1}{2(4 \sigma^2)^2}
C \right]  - \frac{N^2}{\ln 2} \left(\frac{\lambda^2}{4 \sigma^2}
\right)^2 \frac{p - 1}{p^2}.
\label{des}
\end{equation}
Here,
\begin{eqnarray}
C &=& \frac{2 \lambda^4}{p}
- 2 \Lambda_1 \left(1 - \frac{2}{p} + \frac{2}{p^2} \right) -
\nonumber \\
& & 4 \Lambda_2 \frac{(p - 2)}{p} \left( \frac{2}{p} - 1 \right)
- 2 \Lambda_3 \frac{(p - 2)(p - 3)}{p^2}
\end{eqnarray}
with
\begin{eqnarray}
\Lambda_1 &=& \int dr^1 \, dr^2 \, \rho\left(r^1\right) \,
\rho\left(r^2\right) \, \left(r^1 - r^2\right)^4  \\
\Lambda_2 &=& \int dr^1 \, dr^2 \, dr^3 \, \rho\left(r^1\right)
\, \rho\left(r^2\right) \, \rho\left(r^3\right) \,
(r^1 - r^2)^2 (r^1 - r^3)^2 \nonumber \\
\Lambda_3 &=& \int  dr^1 \, dr^2 \, dr^3 \, dr^4 \, \rho\left(r^1\right)
\, \rho\left(r^2\right) \, \rho\left(r^3\right) \, \rho\left(
r^4\right)
(r^1 - r^2)^2 (r^3 - r^4)^2. \nonumber
\end{eqnarray}
Our numerical simulations corroborate that if a quadratic function is fit to the
initial rise of $I(N)$, the coefficients accompanying $N$ and $N^2$ depend
on $p$ and $\sigma$ just as predicted by Eq. (\ref{des}).


\subsection{The limit of vanishing noise}

In the first place, we take $\sigma \to 0$. If
the conditional probability (\ref{ret}) is replaced by a $\delta$-function,
it is readily seen that  $I = \log_2 p$.

In Appendix C we show that for small---but not vanishing---values
of the noise $\sigma$, the mutual information is expected to
grow as
\begin{equation}
\langle I \rangle = \log_2(p) \left[1 - \frac{p - 1}{\log_2p}
\ \left(4 \sqrt{\pi} \sigma B_2 \right)^N \right],
\label{pocoruid}
\end{equation}
where
\begin{equation}
B_2 = \int \rho^2(r) \ dr.
\end{equation}

\begin{figure}
\begin{center}
\rotatebox[origin=tr]{270}{\scalebox{.2}{\includegraphics{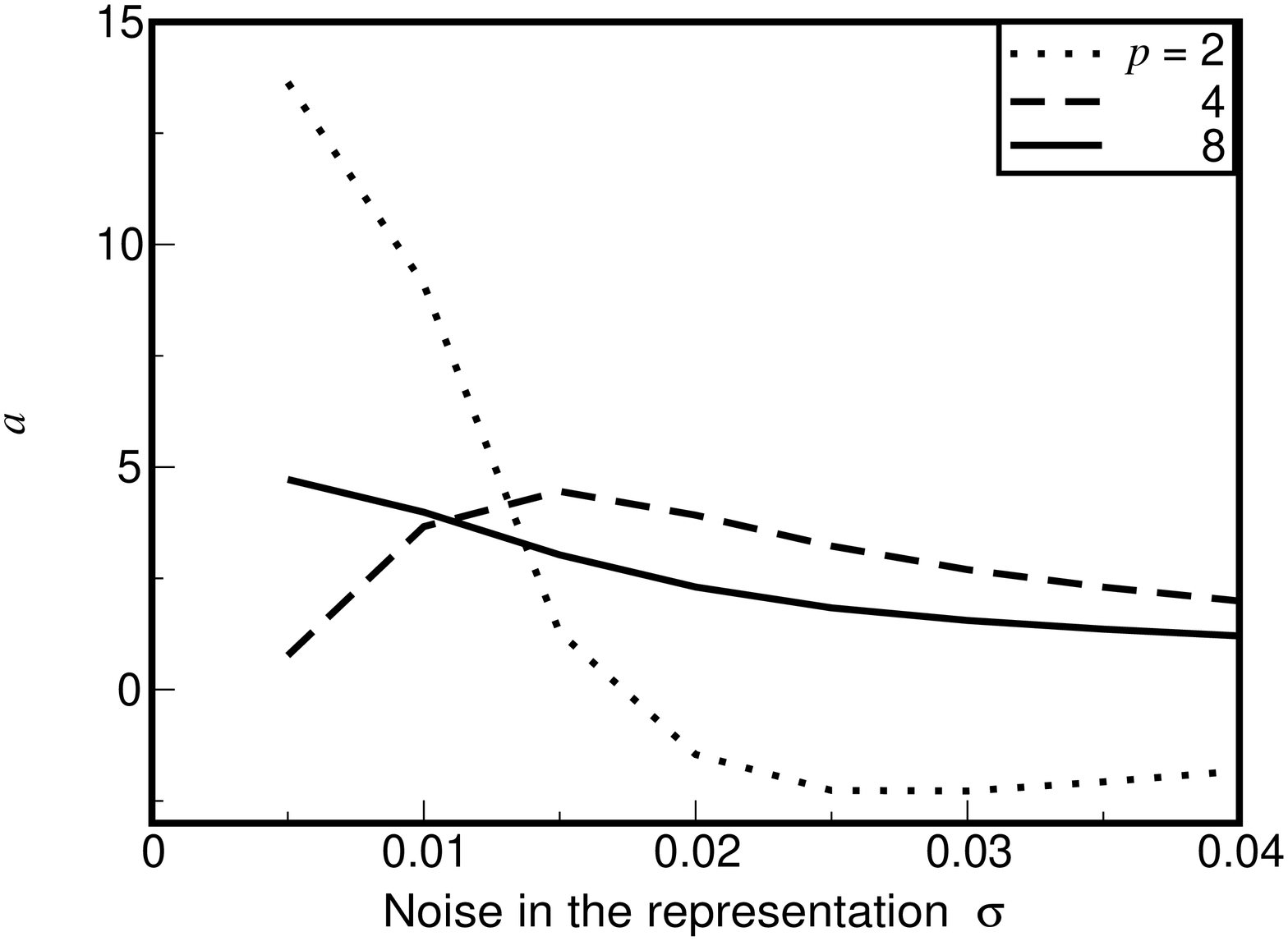}}}
\end{center}
\begin{center}
\rotatebox[origin=tr]{270}{\scalebox{.2}{\includegraphics{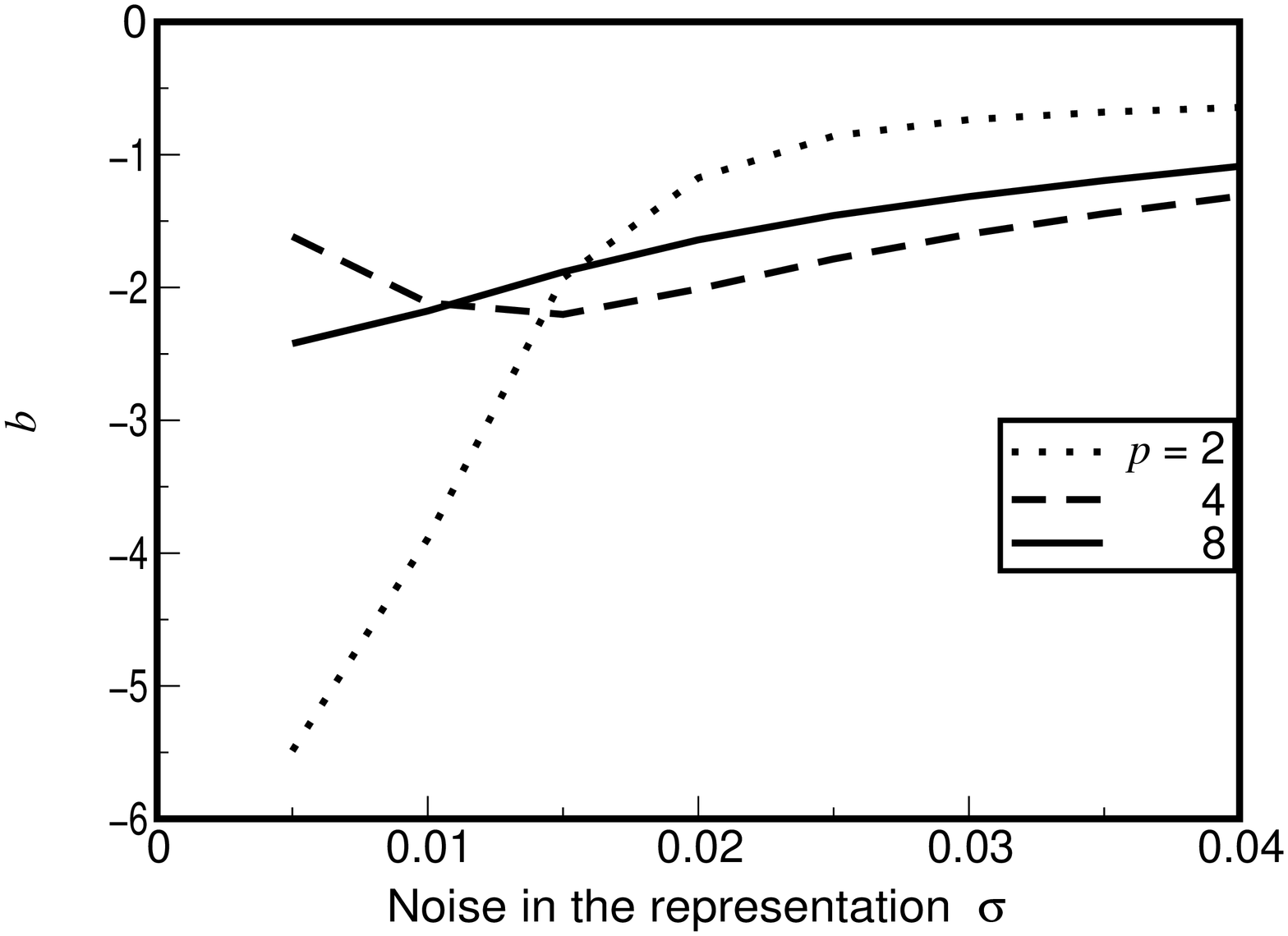}}}
\end{center}
\caption{Dependence of the coefficients $a$ and $b$ a extracted
from numerical evaluations of the mutual information, with the
parameters $p$ and $\sigma$.} \label{fnueva}
\end{figure}

In order to corroborate this result, we have fit a function of the
form $\log_2(p) [1 - a \exp(b N)]$ to the numerical evaluation
of Eq. (\ref{i2e}). In Fig. \ref{fnueva}
we show the dependence of $a$ and $b$ with $\sigma$ and $p$. We
observe that coefficient $a$ shows a dependence with the noise
$\sigma$, in contrast to what is predicted by Eq.
(\ref{pocoruid}). It is also in contrast to the prediction of the
phenomenological model leading to Eq. (\ref{grra}), where $a = 1$.
In addition, $b$ shows a variation with the number of stimuli $p$.
Thus, although it is very easy to calculate the mutual information
when $\sigma$ is exactly equal to 0, we have not been able to
derive analytically the approach to the $\log_2 p$ limit, as
$\sigma \to 0$.


\section{A Related informational measure of accuracy}

Up until now, we have considered the mutual information of Eq.(\ref{inf}), a
quantifier of the capacity with which a given group of units can
represent a fixed set of $p$ stimuli. This is a measure of direct
relevance to neuronal recording experiments. A somewhat different
information measure has been used in analysing mathematical network
models, in particular models of memory storage and retrieval. We would
like to clarify the relationship between the two measures.

Consider the variability with which a typical stimulus is represented,
which in a mathematical model might be described by a formula as simple
as Eq.(\ref{ret}). There, ${\bf r}$ is the response during a trial,
while ${\bf r}^s$ is the average
response across trials with the same stimulus. The average variability
may be quantified by the mutual information between ${\bf r}$ and
${\bf r}^s$,
\begin{equation}
\tilde{I} = \int d{\bf r}^s \ P({\bf r}^s) \int d{\bf r} \ P({\bf r} |
{\bf r}^s) \, \log_2 \left[ \frac{P({\bf r}|{\bf r}^s)}{P({\bf r})}
\right].
\label{iha}
\end{equation}
where ${\bf r}^s$ is taken to span the space of average responses, described by
the probability distribution $P({\bf r}^s)$. In a different model, ${\bf r}^s$ might
be the first response produced, and ${\bf r}$ the second, or any successive response;
in yet other models \cite{ale1,ale2,ale3,Schulz},
${\bf r}^s$ might be the stored representation of a memory
item, and ${\bf r}$ the representation emerging when the item is being retrieved.
In all such cases, one need not refer to a discrete set of $p$ stimuli,
but only to a probability distribution $P({\bf r}^s)$ (and, of course, to a conditional
probability distribution $P({\bf r}|{\bf r}^s)$ ). This measure of accuracy is simply
related to the mutual information we have considered in this paper:
it is given by its $p \to \infty$ limit. In particular, the initial linear
rise of $\tilde{I}$ with $N$ is the only regime relevant to the accuracy measure,
which for independent units is always purely linear in $N$.

Let us see this in formulas. Just as before, we assume that $P({\bf r}^s)$ factorizes as
\begin{equation}
P({\bf r}^s) = \Pi_{j = 1}^N \ \ P(r_j^s).
\end{equation}
The equivalent of (\ref{pglob}) is now
\begin{equation}
P({\bf r}) = \int d{\bf r}^s \, P({\bf r}^s) \, P({\bf r} |
{\bf r}^s).
\label{pedey}
\end{equation}
In Appendix D we show that
\begin{equation}
\tilde{I} = \frac{N}{\ln 2} \frac{\lambda^2}{4 \sigma^2}.
\label{ih}
\end{equation}
In the derivation of Eq. (\ref{ih}) no assumption of small
$N$ has been made. By comparison with Eq. (\ref{imchico}) we
see that,  indeed, the information measure (\ref{iha}) introduced in this
section coincides with the initial rise of the information about
which stimulus is being shown (Sect. III), when the latter
is calculated for a large number of stimuli.


\section{Summary and discussion}

The capacity with which a system of $N$ independent units can
code for a set of $p$ stimuli has been studied. More precisely,
the growth of the mutual information $I$ between stimuli and
responses has been calculated, for different models of the
neural responses. In all these models, the units were supposed
to operate independently. That is to say, the conditional
probability of response ${\bf r}$ given stimulus $s$ is
always a product of single-unit conditional probabilities.
Of course, the fact that neurons operate independently does
not mean that they provide independent information. As
stated in Eq. (\ref{separ}), the mutual information can always
be separated into the difference between the entropy of the responses
($H_1$) and the averaged stimulus specific entropy ($H_2$),
sometimes called noise entropy. For independent units,
$H_2$ is always linear in $N$. However, the factorization
of the conditional probabilities does not imply the factorization
of $P({\bf r})$, meaning that $H_1$ need not be linear in
the number of units. In other words, even independent units
may produce correlated
responses, and indeed strongly correlated, simply because every
unit is driven by the same set of stimuli. Imagine that each
unit provides a very precise representation of the stimuli.
If stimulus 1 is shown, the responses of the $N$ units
will show almost no trial to trial variability.
When the stimulus is changed, another set of $N$ responses
is obtained. But the first responses always come together
(driven by stimulus 1), and so do the second ones.
Even after averaging over all stimuli, this coherent behavior
implies strong correlations between the responses.
In this example,  $H_2 \approx 0$ and $H_1 \approx \log_2 p$.

In other situations, when the number of stimuli is very large,
or the representation of each one of them is noisier, the correlations
in the responses are weaker. We have seen that, in these cases,
$H_1$ tends to become linear in $N$.

Throughout the work, the responses of the units was described
by a vector ${\bf r}$. Nothing was said, however, about what the
components of the vector really are. In the experiment
of Fig. \ref{f1}, $r_j$ was the firing rate of neuron $j$ in a
pre-defined time window. One might however consider a slightly
more complex description in which a subset of $M$ components
is associated to the response of unit $j$. For example, the
first $M$ principal components of its time course
\cite{Optican}. Our analysis would still apply, replacing
$N$ units by $M N$ components.

In the Introduction, reference was made to the
phenomenological models where the growth of $I(N)$
as given by Eq. (\ref{grra}) is entirely explained by
ceiling effects. In such models,
the information provided by different
neurons is supposed to be independent, inasmuch this
is compatible with the fact that the total amount of
information must be $\log_2 p$. The models presented
in this paper are not in principle opposed to the
phenomenological ones; rather they are at a more detailed level
of description. Instead of a direct assumption on how different
units share the available information, we
specify conditional probabilities for the
responses. As a result, we find global trends that
closely resemble those of Eq. (\ref{grra}), that is
to say, an initial linear rise and an exponential
saturation at $\log_2 p$. The detailed shape of
$I(N)$ is, however, different for each model.

It should be kept in mind that whatever the
detailed shape of the curve, the approach to $\log_2 p$
is no more than a consequence of the fact that the
number of stimuli is limited.
The maximum information that can be extracted from the neural
responses is $\log_2 p$. It is clear that if we have a set of
neurons that already provides an information very near to this
maximum, by adding one more neuron we will gain no more than
redundant information. In other words, we have reached a regime
where the neural responses correctly distinguish the identity of
each stimulus. But we cannot deduce from this that the
representational capacity of the responses remains unchanged
when the number of neurons increases. One should rather realize
that the task itself is no longer appropriate to test the way
additional neurons contribute in the encoding of stimuli.
In contrast, the slope of the initial linear rise
is an accurate quantification of the capacity of the system
to represent items.

We have found that distributed coding schemes result
in an initial slope that is roughly independent of
the number of stimuli. This means that the number of
units needed to reach a given fraction of the maximum
information scales as $\log_2 p$---at least, for large
$p$. In contrast, when a grandmother-cell encoding is used,
the initial slope is proportional to $1 / p$, and hence,
one should have $N \propto p$. This makes distributed
encoding much more efficient than localized
schemes.  In the example of the experiment of Fig. \ref{f1},
the information measure supports the conclusion, already
evident from the responses themselves, that the representation
of faces in the inferior temporal cortex of the macaque
is distributed.

\acknowledgments

We thank Damian Zanette for a critical reading of the manuscript.
This work has been supported with a grant of
the Human Frontier Science Programm, number RG 01101998B.

\appendix

\section{Calculation of $H_1$ using the replica method}

Replacing the identity
\begin{equation}
\ln \alpha = \lim_{n \to 0} \frac{1}{n} (\alpha ^n - 1)
\label{rep}
\end{equation}
in (\ref{i2}) the integral in ${\bf r}$ can be evaluated. This
we show in the present appendix.
\begin{eqnarray}
H_1 &=& - \sum_{s = 1}^p  \frac{1}{p}
\int d {\bf r} \ P\left({\bf r} | s \right) \ \frac{1}{\ln 2} \
\lim_{n \to 0} \ \frac{1}{n} \left\{ \left[\sum_{s' = 1}^p \frac{1}{p}
P\left({\bf r} | s' \right) \right]^n - 1 \right\} \nonumber \\
 &=& \frac{-1}{\ln 2} \ \lim_{n \to 0} \frac{1}{n} \ \left(
\frac{H_{1a}}{p^{n + 1}} - 1 \right),
\end{eqnarray}
where
\begin{equation}
H_{1a} = \sum_{s = 1}^p \int d {\bf r} P\left({\bf r} | s \right)
\left[\sum_{s' = 1}^p P \left({\bf r} | s' \right) \right]^n
\sum_{s_1 = 1}^p ... \sum_{s_{n + 1} = 1}^p \Pi_{k = 1}^{N} \ \
H_{1b}(j),
\label{i2a}
\end{equation}
and
\begin{equation}
H_{1b}(j) = \frac{1}{(2 \pi \sigma^2)^{(n + 1) / 2}}
\int d r_j \exp\left[-\frac{1}{2 \sigma^2}
\sum_{k = 1}^{n + 1}
(r_j - r_j^{s_k})^2\right]
\label{inty}
\end{equation}
is a factor that depends on the $j$-th component of one particular
way of distributing $p$ stimuli among the $n + 1$ replicas.
To calculate it we observe that
\begin{equation}
\sum_{k = 1}^{n + 1}(r_j - r^{s_k}_j)^2 = (n + 1) \left[r_j -
\frac{1}{n + 1} \sum_{\ell = 1}^{n + 1} r^{s_\ell}_j \right]^2 +
\vec{\xi}_j^{ \ \dag} \ A \ \vec{\xi}_j,
\label{grando}
\end{equation}
where $\vec{\xi}_j$ is a vector of $n + 1$ components such that
$\xi_j^k = r_j^{s_k}$. The vector notation is used for
arrays of $n + 1$ components. The matrix $A$
has dimensions $(n+1) \times (n+1)$, and reads
\begin{equation}
A = I_d - \frac{1}{n + 1} U,
\label{a}
\end{equation}
where $U$ is an $(n + 1)\times(n + 1)$ matrix, whith all
its coefficients equal to unity.

Thus, the quadratic factor in (\ref{grando}) can be extracted outside
the integral in (\ref{inty}), and
\begin{equation}
H_{1b}(j) = \left(2 \pi \sigma^2 \right)^{-(n + 1)/2} \
\sqrt{\frac{2 \pi \sigma^2}{n + 1}} \
\exp \left(\frac{-1}{2 \sigma^2}
\vec{\xi}_j^{ \ \dag} \ A \ \vec{\xi}_j \right).
\end{equation}
Replacing this expression in (\ref{i2a})
\begin{equation}
H_{1a} = \left[\sqrt{n + 1} (2 \pi \sigma^2)^{n / 2} \right]^{-N} \
\sum_{s_1 = 1}^p ... \sum_{s_{n + 1} =  1}^p \exp\left(
\frac{-1}{2\sigma^2} \sum_{j = 1}^N \vec{\xi}_j^{ \ \dag} \, A \,
\vec{\xi}_j\right)
\label{i2aya}
\end{equation}

We now re-arrange the summation in (\ref{i2aya}), according to the
number $d$ of {\it different} stimuli appearing in the $n + 1$
replicas. For each realization of $s_1, s_2, ... s_{n + 1}$, the
replicas can be divided in $d$ classes, such that all the replicas
belonging to the same class are associated to the same stimulus, and replicas of
different classes correspond to different stimuli.  The number of
replicas adscribed to stimulus $j$ is $K_j$.
Clearly, the sum of all the $K_j$ is $n + 1$, and
only $d$ of the $K_i$ are different from zero. Therefore,
\begin{equation}
\sum_{s_1 = 1}^p \sum_{s_2 = 1}^p ... \sum_{s_{n + 1} = 1}^p =
\sum_{\{K\}}
\left(
\begin{array}{c}
n + 1 \\
\{ K \}
\end{array}
\right).
\label{suma}
\end{equation}
where the term in brackets is defined in (\ref{multin}),
and the $(n + 1)$-fold summation involves all
possible sets of $K_1, ... K_p$
ranging from $0$ to $n + 1$, and whose total sum is $n + 1$.

The advantage of this rearrangement is that the exponent in
(\ref{i2aya}) can be written as a function of only the
differences between representations,
namely
\begin{equation}
\vec{\xi}_j^{ \ \dag} \,  A \, \vec{\xi}_j =
\frac{1}{2(n + 1)} \sum_{m = 1}^p \sum_{\ell = 1}^p
K_m K_\ell (r^m_j - r^\ell_j)^2
\label{expo}
\end{equation}
Therefore, replacing equations (\ref{suma}) and (\ref{expo}) in
(\ref{i2a}) we arrive at Eq (\ref{i2fin})


\section{Initial rise of $\langle I(N) \rangle$ in the large noise limit}
Replacing (\ref{smals}) in (\ref{i2d}), we get
\begin{equation}
\langle H_1 \rangle = -\frac{1}{\ln 2} \ \lim_{n \to 0} \frac{1}{n}
 \left\{
\frac{1}{(n+1)^{M/2} (2 \pi \sigma^2)^{N n / 2} p^{n + 1}}
 \ \left[ p^{n + 1} - \frac{\lambda^2 N}{4(n + 1)\sigma^2}
\ \ S \right]  - 1 \right\},
\label{lio}
\end{equation}
with
\begin{equation}
S =\sum_{\{K\}}
\left(
\begin{array}{c}
n + 1 \\
\{ K \}
\end{array}
\right)
 \ \
\sum_{m = 1}^p \ \ \ \sum_
{
\ell = 1,
\ell \ne m
}^p
K_m \ K_\ell .
\end{equation}
In order to compute $S$ we interchange the order of summation
\begin{equation}
S =
\sum_{m = 1}^p \ \ \ \sum_
{\ell = 1, \ell \ne m }^p
\sum_{\{K\}}
\left(
\begin{array}{c}
n + 1 \\
\{K\}
\end{array}
\right)
 \ \
K_m \ K_\ell .
\label{s}
\end{equation}
The terms with $K_\ell$ or $K_m$ equal to zero, do not contribute to $S$.
Therefore, we can restrict the sum in (\ref{s}) to $K_m \ne 0 \ne
K_\ell$. Thus, the addition over all $K$'s ranging from 0 to $n + 1$ whose
total sum is $n + 1$ can be replaced by another addition, where all
$K$'s different from $K_\ell$ and $K_m$ range from 0 to $n - 1$, $K_m$
and $K_\ell$ go from 1 to $n$, and the sum of all the $K$'s is $n + 1$.
Since there are $p(p - 1)$ choices for $K_\ell$ and $K_m$,
\begin{equation}
S = p(p - 1)(n + 1)n p^{n - 1} \label{ecochi}
\end{equation}
Replacing equation (\ref{ecochi}) in (\ref{lio})
we get
\begin{equation}
\langle H_1 \rangle =
\frac{1}{\ln 2} \left[\frac{N}{2} + \frac{N}{2} \ln(2 \pi \sigma^2) \right]
+  \frac{N}{\ln 2} \frac{p - 1}{p} \left(\frac{\lambda}{2 \sigma}\right)^2.
\end{equation}
When $H_2$ is summed to $\langle H_1 \rangle$, equation
(\ref{imchico}) is obtained.


\section{The small $\sigma$ limit}

We go back to Eq. (\ref{i2d}). We re-write (\ref{i2c}) as
\begin{equation}
\langle A_{\{K\}} \rangle = \int \Pi_{s = 1}^p \ \ d r^s \, \rho\left(
r^s\right)  \ \exp \left[ -\frac{1}{2(n + 1)\sigma^2}
\ \bar{\chi}^{ \ \dag} M \bar{\chi} \right],
\label{intexp}
\end{equation}
where $\bar{\chi}$ is a vector of $p$ components, such that $\chi_s = r^s$,
and
\begin{equation}
M = (n + 1) \left(
\begin{array}{cccc}
K_1 & 0 & ... & 0 \\
0 & K_2 & ... & 0 \\
& & ... & \\
0 & 0 & ... & K_p
\end{array}
\right)
-
\left(
\begin{array}{cccc}
K_1 K_1 & K_1 K_2 & ... & K_1 K_p \\
K_2 K_1 & K_2 K_2 & ... & K_2 K_p \\
& & ... & \\
K_p K_1 & K_p K_2 & ... & K_p K_p
\end{array}
\right).
\end{equation}
The integrand in (\ref{intexp}) is 1 in the origin, and also along the
eignenvectors of $M$ corresponding to a zero eigenvalue. The number of
such eigenvalues is equal or larger than the number of $K$ that are
zero. We therefore re-arrange the numbering of the representations in such a
way as to put all those with $K$ different from zero in the first
$d$ places. Thus, $K_{d + 1} = K_{d + 2} = ...  = K_{p} = 0$. With
this ordering, matrix $M$ is filled with zeros in all those positions
with a row or a column greater than $d$.  Integrating in
$r^{d + 1}, r^{d + 2}, ..., r^{p}$ we get
\begin{equation}
\langle A_{\{K\}} \rangle = \int \Pi_{s = 1}^d \ \ d r^s \, \rho\left(
r^s\right) \ \exp \left[ -\frac{1}{2(n + 1)\sigma^2}
\ \bar{\chi'}^{ \ \dag} M' \bar{\chi'} \right],
\label{intexp1}
\end{equation}
where $\bar{\chi'}$ and $M'$ are defined as $\bar{\chi}$ and $M$, but
live in a space of $d$ dimensions (and not $p$).

In order to integrate Eq. (\ref{intexp1}) we observe that $M'$ has
a single eignevalue $\lambda_1$ equal to zero, with eigenvector
\begin{equation}
w_1 = \frac{1}{\sqrt{d}} \left(
\begin{array}{c}
1 \\
1 \\
... \\
1
\end{array}
\right).
\end{equation}
We call $w_2, ..., w_d$ all the other eigenvectors corresponding to
nonvanishing eigenvalues $\lambda_2, ..., \lambda_d$. We choose
the eigenvectors normalized, and orthogonal to each other and to $w_1$
(the symmety of $M$ allows us to do so).
With this set of vectors we construct a
new basis, and call $\bar{w}$ the collection of coordinates in this
new system. We define a matrix $C$ as the change of basis
\begin{equation}
\bar{\chi} = C \bar{w},
\end{equation}
where
\begin{equation}
C =
\left(
\begin{array}{cccc}
1/\sqrt{d} & c_{12} & ... & c_{1d} \\
1/\sqrt{d}& c_{22} & ... & c_{2d} \\
& & ... & \\
1/\sqrt{d} & c_{d2} & ... & c_{dd}
\end{array}
\right)
\end{equation}
and det($C$) = 1. In this new basis,
\begin{equation}
\langle A_{\{K\}} \rangle = \int \Pi_{j = 1}^d \ \ dw_j
\rho(w_1/\sqrt{d} + c_{j2} w_2 + ... + c_{jd} w_d)
\ \exp\left[-\frac{1}{2} \sum_{\ell = 1}^d \frac{\lambda_\ell}{\sigma^2(n + 1)}
w_\ell^2\right].
\end{equation}
Multiplying and dividing by the product of all $2 \pi (n + 1) \sigma^2 / \lambda_\ell$,
for $\ell \in [2, d]$, we get
\begin{eqnarray}
\langle A_{\{K\}} \rangle &=&
\left(\Pi_{j = 2}^d \  \sqrt{\frac{2 \pi \sigma^2 (n + 1)}
{\lambda_j}} \right) \label{bbl} \\
& & \times
\int \Pi_{j = 1}^d \  \left[dw_j \rho(w_1/\sqrt{d} +
c_{j2} w_2 + ... + c_{jd} w_d)\right] \
\frac{\exp\left[-\frac{1}{2}
\sum_{k = 1}^d \frac{\lambda_k}{\sigma^2(n + 1)}
w_k^2\right]}{\Pi_{j = 2}^d \ \sqrt{\frac{2 \pi \sigma^2 (n + 1)}
{\lambda_j}}}. \nonumber
\end{eqnarray}
In the limit $\sigma \to 0$, the integrand in (\ref{bbl}) includes
$d - 1$ delta functions. Once integrated,
\begin{equation}
\lim_{\sigma \to 0} \langle A_{\{K\}} \rangle = \Pi_{j = 2}^d \
\sqrt{\frac{2 \pi \sigma^2(n + 1)}{\lambda_j}} \ \int dw_1 \,
\left[\rho(w_1/\sqrt{d})\right]^d.
\end{equation}
It may be shown that
\begin{equation}
\Pi_{j = 2}^d  \ \ \lambda_j = d (n + 1)^{d - 2} \Pi_{\ell = 1}^d
\ \ K_\ell.
\end{equation}
Thus,
\begin{equation}
\lim_{\sigma \to 0} \langle A_{\{K\}} \rangle = (2 \pi \sigma^2)^{(d - 1) / 2}
\frac{(n + 1)^{1/2}}{\sqrt{d \ \Pi_{\ell = 1}^d \ K_\ell}} B_d,
\label{mish}
\end{equation}
where
\begin{equation}
B_d = \int dx \left[\rho(x)\right]^d.
\end{equation}

We now turn to the calculation of
\begin{equation}
S = \sum_{\{K\}} \left(
\begin{array}{c}
n + 1 \\
\{ K \}
\end{array} \right) \
\langle A_{\{K\}} \rangle ^N,
\label{sha}
\end{equation}
where, as before, the summation runs over all sets of $\{ K \}$
that add up to $n + 1$. Equation (\ref{mish})
states that $\langle A_{\{K\}} \rangle$ depends on $d$, that is, on
the number of $K$ that are different from zero. Therefore, we write
the sum in (\ref{sha}) as
\begin{equation}
S =
\sum_{d = 1}^{n + 1}\frac{1}{d!}\frac{p!}{(p - d)!}
\ \left[ (2 \pi \sigma^2)^{(d - 1) / 2} \, \sqrt{n + 1}
\, B_d\right]^N
\ \ S_1,
\end{equation}
where
\begin{equation}
S_1 =
\sum_{\{K\}'}
\left(
\begin{array}{c}
n + 1 \\
\{K\} \end{array}
\right) \left[
\frac{1}{\Pi_{j = 1}^d K_j} \right]^{N/2}.
\label{ejuj}
\end{equation}
The sum in $S_1$ involves only the $d$ values of $K$ that are different from
zero. We now make the approximation
\begin{equation}
S_1 \approx \left( \frac{d}{n + 1}\right)^{N d / 2} \
\sum_{\{K\}'}
\left(
\begin{array}{c}
n + 1 \\
\{ K \}
\end{array}
\right).
\end{equation}
But
\begin{equation}
\sum_{\{K\}'}
\left(
\begin{array}{c}
n + 1 \\
\{ K \}
\end{array}
\right) =
\sum_{j = 0}^d (-1)^j \, \frac{d!}{d!(d  - j)!} \, (d - j)^{n + 1}.
\end{equation}
And, taking the limit
\begin{equation}
\lim_{n \to 0}\sum_{\{K\}'}
\left(
\begin{array}{c}
n + 1 \\
\{ K \}
\end{array}
\right) =
\delta_{d, 1} + n \sum_{j = 0}^d \frac{d!}{(d - j)! j!} \ j\ln j \
(-1)^{d - j}.
\end{equation}
Moreover, if $N$ is large, as $d$ grows $(B_d)^N \to 0$. Therefore,
keeping just $d = 1$ and $d = 2$ we may approximate
\begin{equation}
S \approx p + p(p - 1) \, \ln 2 \, (2 \pi \sigma^2)^{N/2} \,
(n + 1)^{N / 2} \, (B_2)^N \ n.
\end{equation}
Replacing in Eqs. (\ref{i2d}) and (\ref{separ}) we arrive at
(\ref{pocoruid}).


\section{Information between the actual response and the stored
representation}

The aim is to calculate (\ref{iha}) under the assumption (\ref{ret}).
Replacing (\ref{ret}) in (\ref{pedey}) the probability $P({\bf r})$
can be written as
\begin{equation}
P({\bf r}) = \Pi_{j = 1}^{N} \ \ \zeta(r_j),
\end{equation}
where
\begin{equation}
\zeta(r_j) = \int d r^0_j \ P(r^0_j) \, \frac{{\rm e}^{
-(r_j - r^0_j)^2 / 2 \sigma^2}}{\sqrt{2 \pi \sigma^2}}.
\label{zeta}
\end{equation}
Just as before, we separate
\begin{equation}
\tilde{I} = H_1 - H_2,
\end{equation}
where
\begin{eqnarray}
H_2 &=& - \int d {\bf r}^0 \int d {\bf r} \ P({\bf r} | {\bf r}^0) P({\bf r}^0)
\log_2\left[P({\bf r} | {\bf r}^0) \right] =
 \frac{N}{2 \ln 2} \left[1 + \ln(2 \pi \sigma^2) \right], \\
H_1 &=& - \int d {\bf r}^0 \int d {\bf r} \ P({\bf r} | {\bf r}^0) P({\bf r}^0)
\log_2\left[P({\bf r})\right] = - N \int d t \ \zeta(t) \ \log_2\left[\zeta(t)\right]
 \nonumber
\end{eqnarray}
Inserting the definition (\ref{zeta}) of $\zeta(t)$, and using the expression
(\ref{rep}) for the logarithm we get
\begin{equation}
H_1 = - \frac{N}{\ln 2} \lim_{n \to 0} \frac{1}{n}
\int d t \ \Pi_{j = 1}^{n + 1} \frac{{\rm e}^{-(t - x_j)^2 /
2 \sigma^2}}{\sqrt{2 \pi \sigma^2}}  -
 \int dx P(x) \int d t \frac{{\rm e}^{-(x - t)^2 / 2 \sigma^2}}
{\sqrt{2 \pi \sigma^2}}. \label{aca}
\end{equation}
The last term in (\ref{aca}) is nothing but the integral of $\zeta(x)$
over all $x$, which can be shown to give 1. To carry out the integral
in $t$ in the first line of (\ref{aca}) we observe that
\begin{eqnarray}
\Pi_{j = 1}^{n + 1} \ \ \frac{{\rm e}^{-(t - x_j)^2 / 2 \sigma^2}}
{\sqrt{2 \pi \sigma^2}}
&=& \left( 2 \pi \sigma^2 \right)^{-(n + 1)/2} \times \nonumber \\
& &\times \exp \left\{\frac{-(n + 1)}{2 \sigma^2} \left[t -
\frac{1}{n + 1} \sum_{j = 1}^{n + 1} x_j\right]^2
- \frac{1}{2 \sigma^2} \left[
\sum_{j = 0}^{n + 1} x_j^2 - \frac{1}{n + 1}\left(
\sum_{k = 1}^{n + 1} x_j \right)^2 \right] \right\} \nonumber
\end{eqnarray}
When replacing this expression in (\ref{aca}), the integration in $t$
can be done right away. The result is $[2 \pi \sigma^2 / (n + 1)]^{1/2}$.
Therefore,
\begin{eqnarray}
H_1 &=& -\frac{N}{\ln 2} \lim_{n \to 0} \frac{1}{n}
\left[ (2 \pi \sigma^2)^{-n/2} \ (n + 1)^{- 1 / 2} \int
\Pi_{\ell = 1}^{n + 1} \ \ d x_\ell P(x_\ell)  \right.
\nonumber \\
& & \left.\exp\left\{\frac{-1}{2 \sigma^2} \left[ \sum_{j = 1}^{n + 1}
x_j^2  - \frac{1}{n + 1} \left(\sum_{k = 1}^{n + 1} x_k \right)^2
\right] \right\}  - 1\right]. \label{aca2}
\end{eqnarray}
In the same way as in Eq. (\ref{expo}), we write
\begin{equation}
\sum_{\ell = 1}^{n + 1} x_\ell^2 - \frac{1}{n + 1}
\left(\sum_{j = 1}^{n + 1} x_j \right)^2 = \frac{1}{2(n + 1)}
\sum_{\ell = 1}^{n + 1} \sum_{m = 1}^{n + 1} (x_\ell - x_m)^2.
\label{aca1}
\end{equation}
Thus, replacing Eq. (\ref{aca1}) in (\ref{aca2}), and making the
expansion
\begin{equation}
\exp \left[- \frac{1}{4 \sigma^2 (n + 1)} \sum_{\ell = 1}^{n + 1}
\sum_{m = 1}^{n + 1}(x_\ell - x_m)^2 \right] \approx 1 - \frac{1}{4 \sigma^2(n + 1)}
\sum_{\ell = 1}^{n + 1} \sum_{m = 1}^{n + 1} (x_\ell - x_m)^2,
\label{ximacion}
\end{equation}
$H_1$ can be calculated. The result is
\begin{equation}
H_1 =  \frac{N}{\ln 2} \left\{\frac{1}{2}\left[1 + \ln(2 \pi \sigma^2)
\right] + \frac{\lambda^2}{4 \sigma^2} \right\}.
\end{equation}
When $H_2$ is substracted, Eq. (\ref{ih}) is obtained. It should be
noticed that Eq. (\ref{ximacion}) is not an approximation. The $j$-th order
in the Taylor expansion of the exponential grows as $[n(n + 1)]^j$. Therefore,
only the linear term gives a contribution for $n \to 0$.


\end{document}